\documentstyle[prb,aps,multicol,epsf,eqsecnum]{revtex}
\begin{document}
\draft

\widetext

\title {The problem of phase breaking in the electronic conduction in mesoscopic
systems: a linear-response theory approach}

\author{Pier A. Mello$^{1}$, Yoseph Imry$^{2}$ and Boris Shapiro$^{3}$}
\address{$^{1}$Instituto de F\'{\i}sica, Universidad Nacional Aut\'{o}noma
de M\'{e}xico, 01000 M\'{e}xico D.F., M\'{e}xico} 
\address{$^{2}$Physics Department, Weizmann Institute of Sciences, 
Rehovot, Israel} 
\address{$^{3}$Department of Physics, Technion-Israel Institute of 
Technology, 32000 Haifa, Israel}

\date{\today }

\maketitle
\widetext

\begin{abstract}
We study the problem of electronic conduction in mesoscopic systems when the
electrons are allowed to interact not only with static impurities, but also
with a scatterer (a phase breaker(PB)) that possesses internal degrees of freedom. 
We first analyze the
role of the PB in reducing the coherent interference effects in a
one-electron quantum mechanical system.
In the many-electron system we can make a number of quite general statements
within the framework of linear-response theory and the random-phase approximation. We
cannot calculate the conductivity tensor in full generality: we thus resort
to a model, in which that tensor can be expressed entirely in a
single-electron picture. The resulting zero-temperature conductance can be
written in terms of the total transmission coefficient at the Fermi energy,
containing an additional trace over the states of the PB. 
\end{abstract}

%\pacs{PACS numbers: xxxxxxxxxxx}

\begin{multicols}{2}
\narrowtext

\section{INTRODUCTION}
\label{intro}

The scattering approach to quantum electronic transport 
in mesoscopic systems was devised by Landauer \cite{landauer(phil mag)}
and later extended by a number of other authors (see, for instance, 
as representative articles, Refs. 
\onlinecite{buettiker-imry-landauer-pinhas,buettiker(ibm)} and the references
contained therein). In an independent-electron picture, it aims at
understanding the {\it electric conductance} of a sample in terms of its 
{\it scattering properties}. The problem of electric transport is thus
converted into that of solving the quantum-mechanical scattering problem of
an electron that impinges on the sample through leads that, ideally, extend
to infinity, once the experimental environment the sample is attached to in
the laboratory is disconnected. 
An approach to this problem using the methods of linear-response theory
(LRT) has been given, for instance, by the authors of 
Refs. \onlinecite{levinson,levinson-shapiro} (see also other publications referred to there).

In the original conception of the scattering approach to electronic
transport, inelastic electron scattering or other {\it phase-breaking
mechanisms} are not allowed inside the sample. As a result, the phase of the
wave function is completely coherent in that region. Yet, in various
circumstances the effects of the electron-electron interaction or the
interaction with the phonon field may not be negligible. In a further
development of the theory \cite{buettiker(ibm)}, the single-electron picture
is maintained and phase-breaking events in a given region are modelled by
attaching the system to a ``fake wire'', which in turn is connected to a
phase-randomizing reservoir, so that there is no phase coherence in the wave
function for an electron entering and exiting the reservoir. The chemical
potential of the reservoir is chosen so that the net current along the fake
wire vanishes. This model provides sensible answers and has been used, for
instance, in the study of electric transport through quantum dots \cite
{baranger-mello95,markus model in cavities}.

A number of authors have attempted to generalize the scattering approach to
include inelastic scattering explicitly, instead of modelling it as
described above. In Ref. \onlinecite{feng} the problem of quantum transport in the
presence of phase-breaking scattering is formulated using an exacly soluble
model for the electron-phonon interaction and, using linear-response theory,
a generalized conduction formula is found. Ref. \onlinecite{chen-sorbello} uses
Landauer's approach and analyzes the effect of a single impurity scatterer,
at which both elastic and inelastic processes can occur. The form of the
electron-impurity interaction can be quite general; in the weak-scattering
limit, in which Born approximation is invoked, the authors arrive at a
generalized conduction formula. The possibility of energy exchange with the
scatterer makes Pauli blocking effects important and an extension beyond
Born approximation seems difficult.

The authors of Ref. \onlinecite{stern} analyze in great generality the problem of
quantum-mechanical phase breaking: they consider the interference between
the terms of an electron wave function arising from two different
electron paths and study the effect on that interference of an
``environment'' the electron can interact with. In an experiment in which
{\it the electron and not the environment is measured}, the coordinate of
the latter is integrated upon: as a result, {\it the interference is lost
if, in the two interfering partial waves, the states of the environment are
orthogonal to each other}. It is emphasized that, for this to occur, energy
exchange between the electron and the environment need not be invoked. That
this loss of interference is irretrievable is a quantum-mechanical effect,
common to a number of situations, like the two-slit experiment.

In the present paper we plan to incorporate the mechanism of 
Ref. \onlinecite{stern} --briefly described in the previous paragraph-- to the LRT approach
to transport provided by Refs. \onlinecite{levinson,levinson-shapiro}, in order to
study the problem of phase breaking in the electronic conduction in mesoscopic
systems. The emphasis on the mechanism of Ref. \onlinecite{stern} for phase
breaking is the main difference between the present and previous work on the
problem. Another characteristic of the present paper is that the
applications are discussed in terms of the scattering matrices of the
various impurities (static or non-static), a procedure that has been found
advantageous in a number of previous publications 
(see Refs. \onlinecite{les houches,wavesrm} and references contained therein).

Quite generally, we can pose the problem by thinking that the electrons,
besides suffering elastic collisions with {\it static} scatterers, interact
with a number of scatterers, or {\it phase breakers} (PB), that possess 
{\it internal degrees of freedom} and can live, say, in $m$ possible
quantum-mechanical states altogether. Even in the absence of
the electron-electron Coulomb interaction, the problem is now no longer one of a
single-electron, but is a full many-body problem: one electron is incident
on the PB, this being in some state $\mu $; after the interaction there is a
certain probability to find the PB in state $\nu $, and this is what the
next electron coming in will see. This {\it memory effect}, or, equivalently, 
the electron-electron interaction induced by the PB, gives rise to a
situation similar to the one found in Kondo problem \cite{kondo problem} and
we are bound to find similar complications. 

The paper is organized as follows. In order to become acquainted with the
physical phenomena produced by a PB we first study, in the next section,
the problem of a {\it single electron} scattered by two static impurities
and a PB. We show how the interference terms that occur with the static
impurities alone are affected by the presence of the PB. The discussion
parallels that given in Ref. \onlinecite{stern} and is done in terms of the
scattering matrices of the various impurities.

In Sec. \ref{LRT} we pose the {\it conduction problem} (a many-electron
problem) of an electronic system with static impurities and a PB [but not
subject to a magnetic field, so that we have {\it time-reversal invariance}
(TRI)], from the standpoint of LRT. We show that we can make a number of
quite general statements. However, we reach a point where we are unable to
calculate the {\it conductivity tensor} in full generality: we thus resort
to a simplified, soluble model that we introduce in Sec. \ref{model}.
The conductivity tensor is calculated within that
model and is found to be expressible entirely in terms of a 
{\it single-electron picture}, i.e. in terms of single-electron Green's
functions. It is then shown that the resulting zero-temperature dc 
{\it conductance} can be expressed in terms of a total transmission coefficient
at the Fermi energy $\epsilon _{F}$, but now containing a trace over the $m$
states of the PB, Eq. (\ref{g 2}). 
Within the restrictions of the model, the
result does not depend on the strength of the electron-PB interaction, as
other analysis do. We present
(see discussion at the end of section III around equation (\ref{lin}))
a speculation as to the validity of our main result beyond the assumptions 
of the soluble model, in the strict linear-transport regime and
above the Kondo temperature (which is taken to be extremely low)
associated with the $m$-level PB.

In Sec. \ref{RMT} we set up a random-matrix description of the electron-PB
system, with possible applications to chaotic cavities, in order to
calculate the effect of the PB on the average conductance and its
fluctuations. The limitations of the model become apparent here.

Our conclusions are discussed in Sec. \ref{concl}.

\section{QUANTUM INTERFERENCE IN A ONE-ELECTRON SCATTERING PROBLEM: THE
EFFECT OF A PHASE BREAKER.}
\label{1el + pb}

We analyze in this section the scattering 
problem of a {\it single electron} interacting with a combination of
three scatterers in series: two static ones and a PB in the middle. We show
how the interference terms that would occur with the static scatterers alone
are affected by the presence of the PB.

For simplicity, the problem is treated as a 1D one. It is described by the
Hamiltonian 
\begin{equation}
H=\frac{p^{2}}{2m}+V_{1}(x)+\sum_{\mu ,\nu }\left| \mu \right\rangle V_{\mu
\nu }(x)\left\langle \nu \right| +V_{3}(x).  \label{H 1el}
\end{equation}
In this equation, $V_{1}(x)$ and $V_{3}(x)$ are the potentials arising from
the two static scatterers, and the third term represents the interaction of
the electron with the PB; the $m$ states of the
latter, denoted by $\mu $, $\nu $, are degenerate in energy.
Below, we shall find it convenient to write our states as 
$m$-component ``spinors'', so that $H$ acquires a matrix form, where rows
and columns are associated with the $m$ states of the PB.

Instead of modelling the potentials for the various scatterers, we have
found it advantageous to model their scattering properties through the
corresponding scattering matrices. In this language, the three scatterers
are described by the $S$ matrices $S_{1}$, $S_{2}$ and $S_{3}$,
respectiverly. We write the $S$ matrix $S_{2}$ for the PB as 
\begin{equation}
S_{2}=\left[ 
\begin{array}{cc}
r_{2} & t_{2}^{\prime } \\ 
t_{2} & r_{2}^{\prime }
\end{array}
\right] ,  \label{S PB}
\end{equation}
where the reflection and transmission matrices (for incidence from the left
or from the right, respectively) $r_{2}$, $r_{2}^{\prime }$, $t_{2}$, $%
t_{2}^{\prime }$ are $m$-dimensional, with matrix elements $r_{2}^{\mu \nu }$%
, etc. The $S$ matrices for the elastic scatterers on the left and on the
right of the PB are written, respectively, as 
\begin{equation}
S_{1}=\left[ 
\begin{array}{cc}
r_{1}I_{m} & t_{1}^{\prime }I_{m} \\ 
t_{1}I_{m} & r_{1}^{\prime }I_{m}
\end{array}
\right] ,  \label{Sel1}
\end{equation}
\begin{equation}
S_{3}=\left[ 
\begin{array}{cc}
r_{3}I_{m} & t_{3}^{\prime }I_{m} \\ 
t_{3}I_{m} & r_{3}^{\prime }I_{m}
\end{array}
\right] ,  \label{Sel3}
\end{equation}
where $r_{1}$, ... , $r_{3}$, ... , are just complex numbers (we are in 1D)
and $I_{m}$ is the $m$-dimensional unit matrix in the space of the PB
states: recall that scatterers 1 and 3 do not change the state of the PB.

The total transmission matrix $t$ for the chain of three scatterers in
series is given by

 \begin{equation}  
t=\left( t_{3}I_{m}\right) 
\frac{1}{I_{m}-r_{12}^{\prime }r_{3}}t_{12}.  
%\label{t}
\end{equation}
In this equation, $t_{12}$ is the transmission matrix for the system formed
by $S_{1}$ and $S_{2}$, given by 
\begin{equation}   %***
t_{12}=t_{2}\frac{1}{I_{m}-\left( r_{1}^{\prime }\right) r_{2}}\left(
t_{1}I_{m}\right) .  \label{t12}
\end{equation}
Similarly, $r_{12}^{\prime }$, the reflection matrix for the combination $%
S_{1}$, $S_{2}$, is given by 
\begin{equation}  %***
r_{12}^{\prime }=r_{2}^{\prime }+t_{2}\frac{1}{I_{m}-\left( r_{1}^{\prime
}\right) r_{2}}\left( r_{1}^{\prime }I_{m}\right) t_{2}^{\prime }.
\label{rprime12}
\end{equation}

To be more specific, we now choose $r_{2}=r_{2}^{\prime }=0$ in the matrix $%
S_{2}$ of Eq. (\ref{S PB}) that defines the PB. Thus the PB will not change
the electron momentum; however, we shall see that it {\it reduces}, in the
electronic current, {\it the interference among the multiply reflected paths
occurring between the two elastic scatterers}. As a consequence of this
choice, the matrices $t_{2}$ and $t_{2}^{\prime }$ are $m$-dimensional
unitary matrices 
\begin{equation}
t_{2}t_{2}^{\dagger }=I_{m},\qquad t_{2}^{\prime }t_{2}^{\prime \dagger }=I_{m}. 
\label{t2 unitary}
\end{equation}
Also, $t_{12}$ reduces to 
\begin{equation}
t_{12}=t_{2}t_{1}  \label{t12 1}
\end{equation}
and $r_{12}^{\prime }$ to

\begin{equation}
r_{12}^{\prime }=t_{2}r_{1}^{\prime }t_{2}^{\prime }  \label{rprime12 1}
\end{equation}
We thus have, for $t$%
\[
t=\left( t_{3}t_{1}\right) \frac{1}{I_{m}-\left( r_{1}^{\prime }r_{3}\right)
t_{2}t_{2}^{\prime }}t_{2} 
\]
\begin{equation}
=\left( t_{3}t_{1}\right) t_{2}\frac{1}{I_{m}-\left( r_{1}^{\prime
}r_{3}\right) t_{2}^{\prime }t_{2}}=\left( t_{3}t_{1}\right) t_{2}\frac{1}{%
I_{m}-au},  \label{t}
\end{equation}
where we have defined the complex number 
\begin{equation}
a=r_{1}^{\prime }r_{3}=\left| a\right| \exp (i\rho )  \label{a}
\end{equation}
and the unitary matrix 
\begin{equation}
u=t_{2}^{\prime }t_{2}.  \label{u}
\end{equation}
The total transmission matrix $t$ of Eq. (\ref{t}) is $m$-dimensional. Its
element $t^{\mu \nu }$ gives the probability amplitude for the process:
\{the electron comes from the left, the PB being in state $\nu $\}$%
\longrightarrow $ \{the electron is transmitted to the right, the PB being
shifted to the state $\mu $\}. The corresponding probability is $T^{\mu \nu
}=|t^{\mu \nu }|^{2}$. Now, $T^{\nu }=\sum_{\mu }T^{\mu \nu }$ is the
transmission probability when the PB is initially in %***
 state $
\nu $ and {\it the PB is not observed; }$T^{\nu }$ can be written as 
\[
T^{\nu }=(t^{\dagger }t)^{\nu \nu } 
\]
\begin{equation}
=T_{3}T_{1}\left[ \frac{1}{I_{m}-a^{*}u^{\dagger }}\frac{1}{I_{m}-au}\right]
^{\nu \nu },  \label{Tnu}
\end{equation}
where Eq. (\ref{t2 unitary}) was used. 
Let us emphasize that the superscript $\nu$ specifies the initial
{\it pure} state of the PB. If, however, the PB is initially in a
{\it mixed} state (a situation of particular interest for the
conductance problem, Sec. \ref{calc sigma}), 
then an additional sum over $\nu$
is needed. Assuming that the PB can be found, with equal probability,
in each of its $m$ states, one obtains for the transmission coefficient $T$:
\begin{equation}
T=\frac{1}{m}\sum_{\nu}T^{\nu}=\frac{1}{m}tr(t^{\dagger }t)  \label{T}
\end{equation}
\begin{equation}
=T_{1}T_{3}\frac{1}{m}tr\left[ \frac{1}{I_{m}-a^{*}u^{\dagger }}\frac{1}{%
I_{m}-au}\right] .  \label{T 1}
\end{equation}

It is instructive to expand $t$ of Eq. (\ref{t}) as the power series 
\begin{equation}
t=\left( t_{3}t_{1}\right) t_{2}\left[I_m+au+a^{2}u^{2}+\cdot \cdot \cdot
\right] .  \label{series for t}
\end{equation}
This series can be easily interpreted in terms of the multiple scattering
processes that occur between the two elastic scatterers, influenced by the
PB in the middle. The transmitted wave function is a linear combination of
all these terms, arising from internal multiple reflections.

We now study a number of particular cases.

\subsection{The Case m=1}
\label{1 el m=1}

We set 
\begin{equation}
t_{2}=t_{2}^{\prime }=1.  \label{t2 1}
\end{equation}
From Eq. (\ref{t}) we have 
\begin{equation}
t=\frac{t_{3}t_{1}}{1-r_{1}^{\prime }r_{3}}=\frac{t_{3}t_{1}}{1-a}.
\label{t 1}
\end{equation}
In this case, scatterer 2 is not a PB, but a static scatterer: it is thus
like having just the two elastic scatterers [more generally, we could choose 
$t_{2}=e^{i\alpha },t_{2}^{\prime }=e^{i\beta }$; then the effect of
scatterer 2 would simply be the addition of the relative phase $(\alpha
+\beta )$ between the original scatterers 1 and 3; that extra phase could
equally well be obtained, for instance, by setting the two elastic
scatterers a distance $d$ farther apart, where $kd=(\alpha +\beta )$]. The
transmission probability discussed above is 
\[
T^{1}=T=\frac{T_{3}T_{1}}{\left| 1-a\right| ^{2}}=T_{3}T_{1}\left| \frac{1+a%
}{1-a^{2}}\right| ^{2} 
\]
\begin{equation}
=T_{3}T_{1}\frac{1+\left| a\right| ^{2}+2Rea}{\left| 1-a^{2}\right| ^{2}}%
\equiv T_{coh}.  \label{I m1}
\end{equation}
This result will be termed the {\it fully coherent response} $T_{coh}$.

\subsection{The Case m=2}
\label{2 el m=2}

Now the PB has two (orthogonal) states. We follow the various multiply
scattered terms occurring between the two elastic scatterers --as given by
Eq. (\ref{series for t})-- in order to understand more closely what the PB
does to them. We consider two examples.

1. Let 
\begin{equation}
t_{2}=t_{2}^{\prime }=\sigma _{x}=\left[ 
\begin{array}{cc}
0 & 1 \\ 
1 & 0
\end{array}
\right] ,  \label{t2 2}
\end{equation}
so that, from Eq. (\ref{u}) 
\begin{equation}
u=t_{2}^{\prime }t_{2}=I_{2}=\left[ 
\begin{array}{cc}
1 & 0 \\ 
0 & 1
\end{array}
\right] .  \label{u 2}
\end{equation}
We examine the multiple-scattering series (\ref{series for t}). In each
passage through the PB, the state of the latter is shifted to the orthogonal
state. But, after the pair of reflections described by the product $%
a=r_{1}^{\prime }r_{3}$, the PB is visited twice and is then back to the
original state. In other words, in each passage, the PB exactly undoes what
it did in the previous one. This is the significance of $u=t_{2}^{\prime
}t_{2}=I_{2}$ in Eq. (\ref{u 2}).

We thus find for $t$, Eq. (\ref{t}) 
\begin{equation}
t=\left( t_{3}t_{1}\right) t_{2}\frac{1}{1-r_{1}^{\prime }r_{3}},
\label{t 2}
\end{equation}
which leads to Eq. (\ref{I m1}), exactly as for the case  with no PB.

2. Let 
\begin{equation}
u=t_{2}^{\prime }t_{2}=\left[ 
\begin{array}{cc}
0 & 1 \\ 
1 & 0
\end{array}
\right] =\sigma _{x}.  \label{u 3}
\end{equation}
This could be obtained, for instance, with the choices 
\begin{equation}
t_{2}^{\prime }=-i\sigma _{y}\qquad t_{2}=\sigma _{z},  \label{t2 3}
\end{equation}
or 
\begin{equation}
t_{2}=t_{2}^{\prime }=\frac{1}{2}\left[ 
\begin{array}{cc}
1+i & 1-i \\ 
1-i & 1+i
\end{array}
\right] .  \label{t2 4}
\end{equation}
Now $u$ shifts the two PB states: i.e., state $\left| 1\right\rangle 
$ is shifted to $\left| 2\right\rangle $ and viceversa. This fact has
important consequences. The multiple-scattering series (\ref{series for t})
for $t$ now gives 
\[
t=\left( t_{3}t_{1}\right) t_{2} 
\]
\begin{equation}
\cdot \left[ I_{2}+a\sigma _{x}+a^{2}I_{2}+a^{3}\sigma _{x}+\cdot \cdot
\cdot \right] ,  \label{t 3}
\end{equation}
which divides naturally into an even-order and an odd-order contribution,
which can be summed up to give 
\begin{equation}
t=\left( t_{3}t_{1}\right) t_{2}\left[ \frac{1}{1-a^{2}}I_{2}+\frac{a}{%
1-a^{2}}\sigma _{x}\right] .  \label{t 4}
\end{equation}
Suppose that in the incident state the electron comes from the left and the
PB is in the {\it pure state}
\begin{equation}
\left| 0\right\rangle =\left[ 
\begin{array}{c}
\alpha \\ 
\beta
\end{array}
\right] ,  \label{inc wf}
\end{equation}
say. The transmitted wave function on the right is thus
\[
\left| \Psi _{k}\right\rangle _{trans}^{0}=\frac{t_{3}t_{1}}{2(1-a^{2})}e^{ikx}
\]
\begin{equation}
\times\left[ 
\begin{array}{c}
(1+a)(\alpha +\beta )+(1-a)(\alpha -\beta )i \\ 
(1+a)(\alpha +\beta )-(1-a)(\alpha -\beta )i
\end{array}
\right] .  \label{transm wf}
\end{equation}
The transmission probability, that we call $T^{0}$, is given by 
\begin{equation}
T^{0}=T_{3}T_{1}\frac{1+\left| a\right| ^{2}+2{\rm Re}(\alpha \beta ^{*})2%
{\rm Re}(a)}{\left| 1-a^{2}\right| ^{2}}.  \label{I 2}
\end{equation}
In the absence of the PB, on the other hand, we have the fully coherent
response $T_{coh\text{ }}$of Eq. (\ref{I m1}). The interference term $2Re(a)$
in Eq. (\ref{I m1}) has been reduced, in Eq. (\ref{I 2}), by a factor, whose
magnitude $\left| 2{\rm Re}(\alpha \beta ^{*})\right| \leq 1$. {\it The
effect of the PB, and of having measured the electron but not the PB, has
thus been to decrease the magnitude of the interference term.} We stress
again that this effect is there even if the incident state is the {\it pure}
state (\ref{inc wf}). In particular, for either pure state $\alpha =1$, $%
\beta =0$, or $\alpha =0$, $\beta =1$, we obtain, in the notation introduced
right after Eq. (\ref{u}) 
\begin{equation}
T^{1}=T^{2}=T_{3}T_{1}\frac{1+\left| a\right| ^{2}}{\left| 1-a^{2}\right|
^{2}},  \label{I 3}
\end{equation}
where the interference term in the numerator %***
 is absent. This last result could be obtained
directly from Eq. (\ref{Tnu}). For a {\it mixture} of these two states we
thus obtain the same answer, i.e. $T=T^{1}=T^{2}$.

\subsection{The Case $m\rightarrow \infty $}
\label{1 el m infty}

Let us inquire as to {\it what kind of a PB would lead to the ``classical''
composition rule for the two elastic scatterers}, i.e. to the equation 
\begin{equation}
T=\frac{T_{1}T_{3}}{1-R_{1}R_{3}}.  \label{Tcl}
\end{equation}
We shall see that for this to occur we need a PB with many states, i.e. $%
m\rightarrow \infty $. We analyze two possibilities.

1. Choose 
\begin{equation}
(t_{2})_{\mu \nu }=(t_{2}^{\prime })_{\mu \nu }
=\delta _{\nu ,\mu -1}.  
\label{t2 5}
\end{equation}
For finite $m$ we use periodic boundary conditions, i.e. $m+1\rightarrow 1$.
But below we are interested in $m\rightarrow \infty $. Assume that the
electron is impinging from the left and the PB is initially, say, in its
first state $|1\rangle $. Each passage of the electron through the PB will
flip the latter to its next (orthogonal) state. Therefore, from the
expansion (\ref{series for t}), the transmitted wavefunction is: 
\begin{equation}
\left| \Psi _{k}\right\rangle _{trans}^{1}=e^{ikx}t_{3}t_{1}[|2\rangle
+a|4\rangle +a^{2}|6\rangle +\cdots ]  \label{transm wf 1}
\end{equation}
It is now clear that the transmission coefficient $T^{1}$ for the electron
(without observing the PB) is given by Eq. (\ref{Tcl}). \newline

2. We now discuss a model that allows varying the degree of dephasing in a
continuous fashion, thus permitting the description of the crossover between
the fully coherent response (\ref{I m1}) and the fully incoherent, or
classical, one (\ref{Tcl}).

We choose $t_{2}=t_{2}^{\prime }=exp(iH/2)$ where $H$ is an $m\times m$
Hermitean matrix. Its eigenvalues $\theta $ are chosen so that their
density $mg(\theta )$ has width $\lambda $ (a trivial way to get that is to
take a diagonal matrix and put the required spectrum by hand, then make an
arbitrary unitary transformation). Obviously, the eigenvalues of $t_{2}^{2}$
will be $exp(i\theta )$, where $\theta $ is an eigenvalue of $H$.
Assuming that the PB is in a mixed state and, thus, using Eq. (\ref{T}),
 we have 
\begin{equation}
T=T_{1}T_{3}\frac{1}{m}\sum_{r=1}^{m}\frac{1}{|1-ae^{i\theta _{r}}|^{2}},
%\label{T 1}
\end{equation}
which in the $m\rightarrow \infty $ limit can be written as 
\begin{equation}
T=T_{1}T_{3}\int_{0}^{2\pi }\frac{g(\theta )d\theta }{|1-ae^{i\theta }|^{2}}.  
\label{T 2}
\end{equation}
Rather then attempt to do the last integral exactly for a given 
$g(\theta )$, it is very instructive to expand, as in Eq. (\ref{series for t}), 
each of the two factors in the denominator as a sum of powers such as
$a^{k}\exp (ik\theta )$, and the complex conjugate, 
$\left( a^{*}\right)^{k^{\prime }}\exp (-ik^{\prime }\theta )$. All the ``diagonal'' terms give
the classical result, Eq. (\ref{Tcl}), as above. The mixed terms with $k$ $%
\neq $ $k^{\prime }$ give the interference. With vanishing $\lambda $, we
get the full interference terms. Otherwise the interference terms are killed
once $|k-k^{\prime }|\lambda >O(1)$. Thus $1/\lambda $ plays the role of the
dephasing length, in units of the distance between the static scatterers.

It is possible, using the above, to analyze a number of interesting physical
situations. Consider $a$ real and close to unity, i.e. 
$a=1-\delta $, with $0<\delta \ll 1$ and $g(\theta )$ picked at $\theta =0$.
Take, for example, the peak of the resonant transmission for the
fully coherent case, $\lambda =0$. The height of the resonance is $\sim 1/\delta ^2$.
The relevant number of bounces 
(the number of significant terms in the series obtained upon expansion of 
$1/(1-ae^{i\theta })$ needed to form the resonance)
is $\sim 1/\delta $.
Now introduce a finite $\lambda $. If $\lambda \ll \delta$, it will affect distant
(i.e. further than $1/\delta $) terms in the series and thus will have no effect
on the height of the resonance.
Dephasing starts to ``hurt" the peak once $\lambda \gg \delta$.

\section{THE CONDUCTION PROBLEM IN THE PRESENCE OF A PHASE BREAKER IN
LINEAR-RESPONSE.}
\label{LRT}

\begin{figure}[tbh]
\epsfxsize=7cm
\centerline{\epsffile{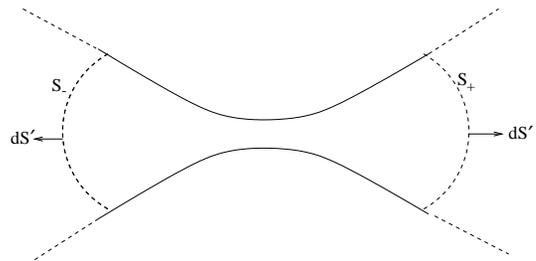}}
\caption{\footnotesize The geometry for the electronic conduction problem
studied in the text. The constriction represents the sample and the
expanding horns, the external leads.
%The regions of integration in Eq. (\protect\ref{J dphi 3}). 
Beyond the surfaces $S_{+}$ and $S_{-}$ the potential 
$\delta \phi ^{\omega }$ takes on constant values.}
\label{horns1}
\end{figure}

In this section we discuss the conduction problem --a
many-electron problem-- in the presence of a PB from the standpoint of
linear-response theory (LRT) \cite{doniach}, following the %***
random phase approximation (RPA) scheme developed
in Refs. \onlinecite{levinson} and \onlinecite{levinson-shapiro}. The system to be
studied has the geometry shown in Fig. \ref{horns1}: the constriction
represents the sample, where we allow for the presence of a PB. As usual,
the expanding horns represent the external leads that, in a laboratory
setup, are attached to macroscopic bodies. 
%\begin{figure}[tbh]
%\begin{center}
%\leavevmode
%\parbox{0.35\textwidth}
%           {\psfig{file=horns.eps,width=0.35\textwidth,angle=0}}
%\end{center}
%\caption{\protect\small The geometry for the electronic conduction problem
%studied in the text. The constriction represents the sample and the
%expanding horns, the external leads.}
%\label{horns}
%\end{figure}

%\begin{figure}
%\narrowtext
%\centerline{\psfig{figure=horns11.ps,width=7cm}}
%\vspace{.5cm}
%\caption{
%The geometry for the electronic conduction problem
%studied in the text. The constriction represents the sample and the
%expanding horns, the external leads.
%The regions of integration in Eq. (\protect\ref{J dphi 3}). 
%Beyond the surfaces $S_{+}$ and $S_{-}$ the potential 
%$\delta \phi ^{\omega }$ takes on constant values.}
%\label{horns1}
%\end{figure}
When no external voltage is applied, the whole system is in equilibrium and
is described by the unperturbed Hamiltonian 
\begin{eqnarray}
H_{0}=\sum_{i}\left[\frac{p_{i}^{2}}{2m}
-e\sum_{I}\frac{e_{I}}{\left| {\bf r}_{i}-{\bf R}_{I}\right| }
-e\sum_{j}\frac{e_{j}}{\left| {\bf r}_{i}-{\bf R}_{j}\right| } \right] \nonumber \\
+\sum_{i<j}\frac{e^{2}}{\left| {\bf r}_{i}-{\bf r}_{j}\right| }
+\sum_{\mu =1}^{m}\left| \mu \right\rangle E_{\mu }\left\langle \mu \right| 
+\sum_{\mu ,\nu =1}^{m}\left| \mu \right\rangle V_{\mu \nu }({\bf r}_{i})
\left\langle \nu \right|  \nonumber \\
\label{H0}
\end{eqnarray}
Here, $-e$ is the electronic charge, ${\bf r}_{i}$ the position variable of
the $i$-th conduction electron, ${\bf R}_{I}$ the (static) position of the $%
I $-th ion with positive charge $e_{I}$ (screened by the bound electrons),
and $e_{j}$ and ${\bf R}_{j}$ the charge and position of the $j$%
-th impurity, considered to be static. Thus $H^{0}$ contains the kinetic
energy, the interaction of the electrons with the ions and the static
impurities, as well as the electron-electron interaction, for which no
approximation is assumed for the time being. It also contains, in the last
two terms of Eq. (\ref{H0}), the intrinsic Hamiltonian of the PB and its
interaction with the electrons,
of the type introduced in the previous section for one electron. We
do not consider a static magnetic field to be present, so that the problem
is time-reversal invariant.

We denote by $\rho _{0}({\bf r})$ and $\phi _{0}({\bf r})$ the equilibrium
charge density and potential, respectively, that satisfy Poisson's equation
\begin{equation}
\nabla ^{2}\phi _{0}({\bf r})=-4\pi \rho _{0}({\bf r}).  
\label{poisson 2}
\end{equation}
Application of an external voltage with frequency $\omega$
will cause a current density $J_{\alpha }^{\omega }({\bf r})$
in the system. It will also lead to a change in the charge density
and in the potential. We denote these changes by $\delta \rho ^{\omega }({\bf r)}$
and $\delta \phi ^{\omega }({\bf r})$ and empasize that these are
full changes, with respect to the equilibrium values $\rho _0 ({\bf r})$
and $\phi _0 ({\bf r})$. In the RPA approximation, which is employed
in the present paper, there is no need to separate the full changes
into external and induced parts. The essence of the RPA is to omit
the electron-electron interactions from the Hamiltonian $H_{0}$
but to use, instead, the full potential $\delta \phi ^{\omega }({\bf r})$
(rather than the external one) in the formulation of LRT. The full
potential is then determined self-consistently from the Poisson
equation. We thus have the three following equations 
\cite{levinson-shapiro}:
\begin{equation}
J_{\alpha }^{\omega }({\bf r})
=\int d^{3}{\bf r}^{\prime }\Gamma _{\alpha }^{\omega }\left( {\bf r},{\bf r}^{\prime }\right) 
\delta \phi ^{\omega }({\bf r}^{\prime }),  \label{J dphi 2}
\end{equation}
\begin{equation}
\delta \rho ^{\omega }({\bf r)=}\int d^{3}{\bf r}^{\prime }\Pi ^{\omega
}({\bf r,r}^{\prime })\delta \phi ^{\omega }({\bf r}^{\prime }),
\label{drho dphi}
\end{equation}
\begin{equation}
\nabla ^{2}\delta \phi ^{\omega }({\bf r)}=-4\pi \delta \rho ^{\omega }({\bf %
r)}  \label{poisson 4}
\end{equation}
with the kernels
\begin{equation}
\Gamma_{\alpha }^{\omega }\left( {\bf r},{\bf r}^{\prime }\right) 
=-\frac{i}{\hbar  }\int_{0}^{\infty }d\tau e^{i\omega \tau -\gamma \tau }
\left\langle \left[ \widetilde{j}_{\alpha }({\bf r},\tau ),
\widetilde{\rho}({\bf r}^{\prime },0)\right] \right\rangle _{00} 
\end{equation}
\begin{equation}
\Pi ^{\omega }({\bf r,r^{\prime }})
=-\frac{i}{\hbar }\int_{0}^{\infty }d\tau e^{i\omega \tau -\gamma \tau }
\Bigl< \Bigl[ \widetilde{\rho }_{el}({\bf r},\tau ),
\widetilde{\rho }_{el}({\bf r}^{\prime },0)\Bigr]\Bigr> _{00},  
\label{Pi00}
\end{equation}
where the limit $\gamma\rightarrow 0$ is implied. The expectation 
values of the interaction picture operators are taken with respect
to the Hamiltonian
\begin{equation}
H_{00}=H_{0}-\sum_{i<j}\frac{e^{2}}{\left| {\bf r}_{i}-{\bf r}_{j}\right| }, 
\label{H00}
\end{equation}
with the electron-electron interaction switched off.

The Poisson equation (\ref{poisson 4}) should be supplemented
by boundary conditions. Sufficiently far inside the horns the local current
density $J_{\alpha }^{\omega }({\bf r})$ becomes vanishangly small,
so that these distant regions remain practically in equilibrium.
This means that well inside the horns $\delta \phi ^{\omega }({\bf r})$
approaches constant values, 
$\delta \phi ^{\omega }({\bf -\infty })$ and $\delta \phi ^{\omega }(+{\bf \infty })$. 
The difference 
\begin{equation}
\delta \phi ^{\omega }(-\infty )-\delta \phi ^{\omega }(+\infty )=V^{\omega }
\label{dphi asympt}
\end{equation}
is the total potential drop on the system sample+horns 
($V^\omega$ generally does not coincide with the external EMF, since some voltage drop
can occur in other parts of the circuit, e.g., near the points where
the external EMF source is connected to the horns). At the internal
boundary of the system one should require zero current density
normal to the boundary
\begin{equation}
J_n({\bf r}_s)=0,   
\label{normal}
\end{equation}
${\bf r}_s$ being a point on the internal boundary.

Solving (\ref{J dphi 2}), (\ref{drho dphi}), (\ref{poisson 4})
with the boundary conditions (\ref{dphi asympt}), (\ref{normal})
would enable one to determine the charge, current density and
field distribution within the system. One could then compute
the total current $I^{\omega }$ and, thus, the conductance
$G^{\omega }=I^{\omega }/V^{\omega }$. This is a formidable problem. A great
simplification, however, occurs in the dc limit, thanks to a result
obtained in Ref. \onlinecite{KSL} that states that, in that limit and for
fixed ${\bf r}$, ${\bf r}^{\prime }$,  the conductivity tensor is
divergenceless; i.e. 
\begin{equation}
\partial _{\beta }^{\prime }\sigma _{\alpha \beta }^{\omega }\left( {\bf r},%
{\bf r}^{\prime }\right) \rightarrow 0 ,  \label{div sigma =0}
\end{equation}

The conductivity tensor relates current density to the electric
field [rather than to the potential, as in equation (\ref{J dphi 2})];
i.e.
\begin{equation}
J_{\alpha }^{\omega }({\bf r})=-\int d^{3}{\bf r}^{\prime }\sigma _{\alpha
\beta }^{\omega }\left( {\bf r},{\bf r}^{\prime }\right) \partial _{\beta
}^{\prime }\delta \phi _{\omega }({\bf r}^{\prime }), 
\label{J dphi 13}
\end{equation}
and is given in terms of the current-current correlation function as
\begin{eqnarray}
\sigma _{\alpha \beta }^{\omega }\left( {\bf r},{\bf r}^{\prime }\right) 
&&=\frac{1}{\hbar \omega }\int_{0}^{\infty }d\tau e^{i\omega \tau -\gamma \tau }
\left\langle \left[ \widetilde{j}_{\alpha }({\bf r},\tau ),
\widetilde{j}_{\beta }({\bf r}^{\prime },0)\right] \right\rangle _{00}  \nonumber \\
&&-\frac{e^{2}n_{0}({\bf r})}{im\omega }
\delta ({\bf r}-{\bf r}^{\prime })\delta _{\alpha \beta },  
\label{sigma int}
\end{eqnarray}
$n_{0}({\bf r})$ being the electron density in equilibrium.
Integrating Eq. (\ref{J dphi 13}) by parts shows that, {\it in the dc limit, the
current density is insensitive to the full potential profile within
the sample and depends only on the total potential drop between
the two distant surfaces well inside the horns} (Fig. \ref{horns1}):
\begin{equation}
J_{\alpha }^{\omega \rightarrow 0}({\bf r})
=-\left[ \delta \phi ^{\omega \rightarrow 0}(+{\bf \infty })\Gamma _{\alpha }^{+}({\bf r})
+\delta \phi ^{\omega \rightarrow 0}(-{\bf \infty })\Gamma _{\alpha }^{-}({\bf r})\right],  
\label{J dphi 4}
\end{equation}
where 
\begin{equation}
\Gamma _{\alpha }^{\pm }({\bf r})
=\int_{S_{\pm }}dS_{\beta }^{\prime }
\sigma _{\alpha \beta }^{\omega \rightarrow 0}\left( {\bf r},{\bf r}^{\prime }\right) .  
\label{Gamma +-}
\end{equation}
This observation paves the way for a derivation of a Landauer
formula from the LRT, and demonstrates that interactions, within
RPA, do not affect the conductance. {\it This conclusion, known for
purely elastic scatterers \cite{levinson-shapiro}, remains valid also in 
presence of the PB}.

For zero temperature, and in the absence of PB, the answer is the well
known one
\begin{equation}
G=\frac{e^{2}}{h}\sum_{ab}\left| t_{ab}\right| ^{2},  
\label{BL}
\end{equation}
where $t$ is the {\it single-particle transmission matrix at the
Fermi energy,} from well inside the left horn to deep inside the right one.

When the system is described by the unperturbed Hamiltonian $H_{00}$ 
of Eq. (\ref{H00}), the
electrons can change the state of the PB through the e-PB interaction
represented by the last term in Eq. (\ref{H0}); the coupling of the electrons with
the PB has the consequence that the former are no longer independent. This
results in a complicated structure for the $N$-electron eigenstates of 
$H_{00}$, and the calculation of 
$\sigma _{\alpha \beta }^{0}\left( {\bf r},{\bf r}^{\prime }\right) $ 
is, in principle, no longer feasible along the lines followed in
the absence of the PB. In the next section we discuss a model for the e-PB
interaction that does lead to a solution along similar lines. One might 
speculate on physical grounds that our final result, equation (\ref{g 2}),
should be valid under the following assumptions. One should first take the temperature 
to be very low, but much larger than the Kondo temperature, $T_K$, due to the
interaction of the PB with the electron gas. Furthermore, one should stay 
in the strict linear response regime, where for a finite distance $L$ between
the two reservoirs, the current satisfies:
\begin{equation}
e/I \gg L/v_F .
\label{lin}
\end{equation}
This assures that the separation in time between consecutive electrons
participating in the transport is so large that the first electron
reaches the downstream reservoir and thermalizes there before the next electron
starts its journey. This can be expected to eliminate the electron-electron 
interaction mediated by the PB, which is a coherent second-order process. 

\section{A SOLUBLE MODEL}

\label{model}

We assume that $H_{00}$ of Eq. (\ref{H00}) has such a structure that, in a
suitable PB basis, labelled by $\sigma $, $\sigma ^{\prime }$ below, it
acquires the diagonal form 
\begin{equation}
\overline{H}_{00}^{\sigma \sigma ^{\prime }}=\left\{ \sum_{i}\left[ \frac{%
p_{i}^{2}}{2m}+U^{\sigma }({\bf r}_{i})\right] +\Delta ^{\sigma }\right\}
\delta _{\sigma \sigma ^{\prime }}.  \label{HD}
\end{equation}
Just as we did in Sec. \ref{1el + pb}, we write the Hamiltonian in matrix
form, where rows and columns, i.e. $\sigma $, $\sigma ^{\prime }$, label the 
$m$ states of the PB. The bar in $\overline{H}_{00}^{\sigma \sigma ^{\prime
}}$indicates that the Hamiltonian is expressed in the new PB basis, that we
shall call, for short, the D-basis (D for diagonal), as opposed to the
original, or ND-basis.

In the ND basis, $H_{00}$ is obtained from $\overline{H}_{00}$ by means of a 
{\it constant}, real (in order to preserve reality of the Hamiltonian, and
hence time-reversal invariance) orthogonal transformation $O$ (an $m$%
-dimensional matrix), i.e. 
\begin{equation}
H_{00}=O\overline{H}_{00}O^{T},  \label{HND(HD)}
\end{equation}
or, in terms of its matrix elements 
\begin{equation}
H_{00}^{\mu \nu }=\sum_{i}\left[ \frac{p_{i}^{2}}{2m}\delta _{\mu \nu
}+\sum_{\sigma =1}^{m}O^{\mu \sigma }U^{\sigma }({\bf r}_{i})O^{\nu \sigma
}\right] +\sum_{\sigma =1}^{m}O^{\mu \sigma }\Delta ^{\sigma }O^{\nu \sigma
}.  \label{HND}
\end{equation}
We choose $\Delta ^{\sigma }$ constant, i.e. independent of $\sigma $ (and
hence we set it equal to zero), so as not to have constant terms in the
off-diagonal matrix elements $\mu \neq \nu $. In the language of Eq. (\ref
{H0}), the energies $E_{\mu }$ of the PB states are degenerate and set equal
to zero.

The Schr\"{o}dinger equation in the D and ND-basis is 
\begin{eqnarray}
\overline{H}_{00}\left| \overline{\Psi }\right\rangle &=&E\left| \overline{%
\Psi }\right\rangle ,  \label{Schr eqn D} \\
H_{00}\left| \Psi \right\rangle &=&E\left| \Psi \right\rangle ,
\label{ Schr eqn ND}
\end{eqnarray}
respectively, with 
\begin{equation}
\left| \Psi \right\rangle =O\left| \overline{\Psi }\right\rangle .
\label{D ND basis}
\end{equation}

In the D-basis, the Hamiltonian $\overline{H}_{00}$ can be written as 
\begin{equation}
\overline{H}_{00}^{\sigma \sigma ^{\prime }}=\sum_{i}\overline{H}%
_{00}^{\sigma \sigma ^{\prime }}(i)=\overline{H}_{00}^{\sigma }\delta
_{\sigma \sigma ^{\prime }},  \label{HD 1}
\end{equation}
with 
\begin{equation}
\overline{H}_{00}^{\sigma \sigma ^{\prime }}(i)=\overline{H}_{00}^{\sigma
}(i)\delta _{\sigma \sigma ^{\prime }}  \label{HD 2}
\end{equation}
and 
\begin{equation}
\overline{H}_{00}^{\sigma }(i)=\frac{p_{i}^{2}}{2m}+U^{\sigma }({\bf r}_{i}).
\label{HD 3}
\end{equation}
In the ND basis, 
\begin{equation}
H_{00}^{\mu \nu }=\sum_{i}H_{00}^{\mu \nu }(i),  \label{HND 1}
\end{equation}
with 
\begin{equation}
H_{00}^{\mu \nu }(i)=\frac{p_{i}^{2}}{2m}\delta _{\mu \nu }+U^{\mu \nu }(%
{\bf r}_{i})  \label{HND 2}
\end{equation}
and 
\begin{equation}
U^{\mu \nu }({\bf r}_{i})=\sum_{\sigma =1}^{m}O^{\mu \sigma }U^{\sigma }(%
{\bf r}_{i})O^{\nu \sigma }.  \label{UND}
\end{equation}

If the matrix $M^{\mu \sigma }=\left[ O^{\mu \sigma }\right] ^{2}$ has
nonzero determinant, we can find the $U^{\sigma }({\bf r}_{i})$'s ($\sigma =$%
1,$\cdot \cdot \cdot $, $m$) that reproduce any given set of diagonal
potentials $U^{\mu \mu }({\bf r}_{i})$'s ($\mu =$1,$\cdot \cdot \cdot $, $m$%
); but then we are left with no freedom to select the off-diagonal
interactions $U^{\mu \nu }({\bf r}_{i})$ ($\mu \neq \nu $), which become
uniquely determined by the $U^{\sigma }({\bf r}_{i})$'s. So, it is clear
that the matrix elements $U^{\mu \nu }({\bf r}_{i})$ of Eq. (\ref{UND}) show
strong correlations among themselves. These correlations make it possible to
find a D-basis in which the Hamiltonian takes the form of Eqs. (\ref{HD 1})-(%
\ref{HD 3}) and the problem breaks up into $m$ independent single-particle
ones: this is the feature that makes the problem soluble. Solving the
Schr\"{o}dinger equation (\ref{Schr eqn D}) thus reduces to solving the $m$
single-particle Schr\"{o}dinger equations

\begin{equation}
\left[ \frac{p_{i}^{2}}{2m}+U^{\sigma }({\bf r}_{i})\right] \psi
_{k}^{\sigma }({\bf r}_{i})=\epsilon _{k}^{\sigma }\psi _{k}^{\sigma }({\bf r%
}_{i}),\qquad \sigma =1,\cdot \cdot \cdot ,m.  \label{Schr eqn psi sigma}
\end{equation}
We shall assume that none of the $U^{\sigma }({\bf r}_{i})$'s admits bound
states.

In the D-basis, the states 
\begin{equation}
\left| \overline{\Psi }_{k}^{\sigma }\right\rangle =\left[ 
\begin{array}{c}
0 \\ 
\vdots \\ 
0 \\ 
\psi _{k}^{\sigma }({\bf r}_{i}) \\ 
0 \\ 
\vdots \\ 
0
\end{array}
\right] ,\qquad \sigma =1,\cdot \cdot \cdot ,m,\quad  \label{sps D}
\end{equation}
with a nonzero value in the $\sigma $-th entry, form a complete set of
orthonormalized (in a $\delta $-function sense) single-particle states,
eigenfunctions of the single-particle Hamiltonian matrix (\ref{HD 2}).

In the D-basis, the $S$ matrix associated with a scattering solution has the
form 
\begin{equation}
\overline{S}^{\sigma \sigma ^{\prime }}=\overline{S}^{\sigma }\delta
_{\sigma \sigma ^{\prime }}.  \label{SD}
\end{equation}
If the problem admits $N$ open spatial channels, each $S^{\sigma }$ is $2N$%
-dimensional. In the ND-basis $S$ takes the form 
\begin{equation}
S^{\mu \nu }=\sum_{\sigma =1}^{m}O^{\mu \sigma }\overline{S}^{\sigma }O^{\nu
\sigma }  \label{SND}
\end{equation}
and is $2mN$-dimensional. The number of independent parameters associated
with each unitary symmetric matrix $\overline{S}^{\sigma }$ is $N(2N+1)$
and is thus $mN(2N+1)$ for the total $\overline{S}$ and hence for $S$. This
makes it clear that the $S$ matrices allowed by our soluble model do not
have the ``generic'' structure used in some of the considerations of Sec. 
\ref{1el + pb}, but have a rather restricted one: in fact, a generic $mN$%
-dimensional $S$-matrix has a larger number, i.e. $mN(2mN+1)$, of
independent parameters.

For two particles, say, we have the states (not antisymmetrized yet), with $%
\sigma =\pm 1$ 
\begin{equation}
\left| \overline{\Psi }_{k_{1}k_{2}}^{\sigma }\right\rangle =\left[ 
\begin{array}{c}
0 \\ 
\vdots \\ 
0 \\ 
\psi _{k_{1}}^{\sigma }({\bf r}_{i})\psi _{k_{2}}^{\sigma }({\bf r}_{i}) \\ 
0 \\ 
\vdots \\ 
0
\end{array}
\right] .  \label{2-p s D }
\end{equation}

To antisymmetrize we use a second-quantization language, so that for $N$
electrons we have the states 
\begin{equation}
\left| \overline{\Psi }_{k_{1}\cdot \cdot \cdot k_{N}}^{\sigma
}\right\rangle =\left[ 
\begin{array}{c}
0 \\ 
\vdots \\ 
0 \\ 
(c_{k_{1}}^{\sigma })^{\dagger }\cdot \cdot \cdot (c_{k_{N}}^{\sigma })^{\dagger } \\ 
0 \\ 
\vdots \\ 
0
\end{array}
\right] \left| 0\right\rangle ,  \label{N-p s D}
\end{equation}
with $(c_{k}^{\sigma })^{\dagger }$ creating one electron in state $\psi
_{k}^{\sigma }({\bf r})$ and $\left| 0\right\rangle $ being the electron
vacuum.

\subsubsection{The conductivity tensor and the conductance}

\label{calc sigma}

The conductivity tensor $\sigma _{\alpha \beta }^{\omega }\left( {\bf r},%
{\bf r}^{\prime }\right) $ is given in Eq. (\ref{sigma int}). The
expectation value occurring in that equation has to be understood as 
\[
\left\langle \left[ \widetilde{j}_{\alpha }({\bf r},\tau ),\widetilde{j}%
_{\beta }({\bf r}^{\prime },0)\right] \right\rangle _{00}
\]
\begin{equation}
=\sum_{{\cal N}M\sigma }P({\cal N}M;\sigma )\left\langle {\cal N}M;\sigma
\left| \left[ \widetilde{j}_{\alpha }({\bf r},\tau ),\widetilde{j}_{\beta }(%
{\bf r}^{\prime },0)\right] \right| {\cal N}M;\sigma \right\rangle .
\label{av(O)}
\end{equation}
Here, $\left| {\cal N};\sigma \right\rangle $ are the states of Eq. (\ref
{N-p s D}) in the D PB basis, ${\cal N}$ being the number of electrons and $M
$ an abbreviation for the configuration $k_{1}$, $k_{2}$, $\cdot \cdot \cdot 
$, $k_{{\cal N}}$. The states $\left| {\cal N}M;\sigma \right\rangle $ are
eigenstates of the Hamiltonian $\overline{H}_{00}$ of Eq. (\ref{HD 1}), with
the energy $E_{NM;\sigma }=\epsilon _{k_{1}}^{\sigma }+\cdot \cdot \cdot
+\epsilon _{k_{N}}^{\sigma }$.

The current operator $\widetilde{j}_{\alpha }({\bf r},\tau )$ in the
interaction representation is given by 
\begin{equation}
\widetilde{j}_{\alpha }({\bf r},\tau )=e^{\frac{i}{\hbar }\overline{H}%
_{00}\tau }j_{\alpha }({\bf r})e^{-\frac{i}{\hbar }\overline{H}_{00}\tau }.
\label{jtilde}
\end{equation}
In the D PB basis the Hamiltonian  $\overline{H}_{00}$ has the diagonal form
given by Eq. (\ref{HD 1})-(\ref{HD 3}); since the current operator $%
j_{\alpha }({\bf r})$ does not depend on the PB explicitly, $\widetilde{j}%
_{\alpha }({\bf r},\tau )$ takes the diagonal form 
\begin{equation}
\widetilde{j}_{\alpha }^{\sigma \sigma ^{\prime }}({\bf r},\tau )=\widetilde{%
j}_{\alpha }^{\sigma }({\bf r},\tau )\delta _{\sigma \sigma ^{\prime }},
\label{jtilde 1}
\end{equation}
where 
\begin{equation}
\widetilde{j}_{\alpha }^{\sigma }({\bf r},\tau )=e^{\frac{i}{\hbar }%
\overline{H}_{00}^{\sigma }\tau }j_{\alpha }({\bf r})e^{-\frac{i}{\hbar }%
\overline{H}_{00}^{\sigma }\tau },  \label{jalpha tilde sigma}
\end{equation}
$\overline{H}_{00}^{\sigma }$ being given by Eq. (\ref{HD 1}).

Definition (\ref{av(O)}) implies, as usual, that the density matrix is
diagonal in the representation in which the Hamiltonian is diagonal, with
diagonal elements $P({\cal N}M;\sigma )$. Explicitly, $P({\cal N}M;\sigma )$
is given by the grand-canonical ensemble (understanding now the labels $%
{\cal N}$, $M$ as the set of occupation numbers $n_{1}$, $n_{2}$, $\cdot
\cdot \cdot $ ) as 
\begin{equation}
P(n_{1},n_{2},...;\sigma )=\frac{e^{-\beta \sum_{i=1}^{\infty
}n_{i}(\epsilon _{i}^{\sigma }-\mu )}}{{\cal Z}(\beta ,\mu )},  \label{c}
\end{equation}
where the grand partition function is 
\begin{equation}
{\cal Z}(\beta ,\mu )=\sum_{\sigma =1}^{m}{\cal Z}(\beta ,\mu ;\sigma ),
\label{d}
\end{equation}
with 
\begin{equation}
{\cal Z}(\beta ,\mu ;\sigma )=\sum_{n_{i}=0,1}e^{-\beta \sum_{i=1}^{\infty
}n_{i}(\epsilon _{i}^{\sigma }-\mu )}.  \label{e}
\end{equation}
We define the conditional occupation probability 
\begin{equation}
P(n_{1},n_{2},...|\sigma )=\frac{e^{-\beta \sum_{i=1}^{\infty
}n_{i}(\epsilon _{i}^{\sigma }-\mu )}}{{\cal Z}(\beta ,\mu ;\sigma )},
\label{f}
\end{equation}
the condition being that the PB be precisely in the state $\sigma $ of the D
basis; its trace is 1. Then 
\begin{equation}
P(n_{1},n_{2},...;\sigma )=p(\sigma )P(n_{1},n_{2},...|\sigma ),  \label{g}
\end{equation}
where 
\begin{equation}
p(\sigma )=\frac{{\cal Z}(\beta ,\mu ;\sigma )}{{\cal Z}(\beta ,\mu )}
\label{h}
\end{equation}
is the probability of finding the PB in state $\sigma $, the temperature and
chemical potential, not indicated explicitly, being $\beta ,\mu $. Of course
we have 
\begin{equation}
\sum_{\sigma =1}^{m}p(\sigma )=1,  \label{i}
\end{equation}
and so the trace of (\ref{g}) is 1.

We can thus write the expectation value (\ref{av(O)}) as 
\begin{eqnarray}
&&\left\langle \left[ \widetilde{j}_{\alpha }({\bf r},\tau ),
\widetilde{j}_{\beta }({\bf r}^{\prime },0)\right] \right\rangle _{00} \nonumber \\
&&\;\;\;=\sum_{\sigma = 1}^{m}p(\sigma )\sum_{n_{1},n_{2},
\cdot \cdot \cdot }P(n_{1},n_{2},...|\sigma ) \nonumber \\
&&\;\;\;\;\;\;\times \left\langle n_{1},n_{2},...;\sigma \left| 
\left[ \widetilde{j}_{\alpha }^{\sigma }({\bf r},\tau ),
\widetilde{j}_{\beta }^{\sigma }({\bf r}^{\prime },0)\right] 
\right| n_{1},n_{2},...;\sigma \right\rangle . \nonumber \\
\label{av(comm)}
\end{eqnarray}

Thus in the D PB basis {\it the problem breaks up into m independent problems%
}, for the Hamiltonians $\overline{H}_{00}^{\sigma }$ , $\sigma =1,\cdot
\cdot \cdot ,m$. The conductivity tensor $\sigma _{\alpha \beta }^{\omega
}\left( {\bf r},{\bf r}^{\prime }\right) $ of Eq. (\ref{sigma int}) can thus
be written as 
\begin{equation}
\sigma _{\alpha \beta }^{\omega }\left( {\bf r},{\bf r}^{\prime }\right)
=\sum_{\sigma =1}^{m}p(\sigma )\sigma _{\alpha \beta }^{\omega ;\sigma
}\left( {\bf r},{\bf r}^{\prime }\right) ,  \label{sigma int 1}
\end{equation}
where $\sigma _{\alpha \beta }^{\omega ;\sigma }\left( {\bf r},{\bf r}%
^{\prime }\right) $ is a conductivity tensor that can be expressed entirely
in terms of single-electron Green's functions \cite{levinson,KSL} for the
Hamiltonian $\overline{H}_{00}^{\sigma }(i)$.

We write the conductance $G$ in terms of the ``dimensionless conductance'' $%
g $ as

\begin{equation}
G=\frac{e^{2}}{h}g.  \label{G(g)}
\end{equation}
As $\omega \rightarrow 0$ and then the temperature $\rightarrow 0$, we find,
for ``spinless electrons'' 
\begin{equation}
g=T=\sum_{\sigma =1}^{m}p(\sigma )tr\left[ (\overline{t}^{\sigma })^{\dagger
}\overline{t}^{\sigma }\right] ,  \label{g 1}
\end{equation}
$\overline{t}^{\sigma }$ being the transmission matrix (an $N\times N$ block
of the matrix $\overline{S}^{\sigma }$ of Eq. (\ref{SD})) arising from the
potential $U^{\sigma }({\bf r})$; the trace in the above equation is over
spatial channels, as usual. 
Should $U^{\sigma }({\bf r})=U({\bf r})$, i.e. independent of $\sigma $ 
(and hence $U^{\mu \nu }({\bf r})=U({\bf r})\delta _{\mu \nu }$ in any basis), 
the above formula would go over into the
standard one. In the ND PB basis we finally find ($a$, $b$ being
spatial-channel indices) 
\[
T=\sum_{\mu \nu \nu ^{\prime }} \sum_{ab} \rho _{PB}^{\nu \mu }\left[
t_{ab}^{\nu ^{\prime }\mu }\right] ^{*}t_{ab}^{\nu ^{\prime }\nu }
\]
\begin{equation}
=Tr\left( \rho _{PB}t^{\dagger }t\right) ,  
\label{g 2}
\end{equation}
the trace being now over the spatial channels and the PB states. We have
defined 
\begin{equation}
\rho _{PB}^{\nu \mu }=\sum_{\sigma }O^{\nu \sigma }p(\sigma )
O^{\mu \sigma }.  
\label{rhoPB}
\end{equation}
Eq. (\ref{g 2}) has the structure of the standard Landauer formula, with an extra
average over the PB states. This is also the structure of Eq. (\ref{T}),
that was obtained in the study of a {\it single} electron interacting with a
PB with equal weights assigned to every PB state.

Should there be circumstances where the various $p(\sigma )$ discussed
above, at zero temperature, were all equal to $1/m$, we could write 
\begin{equation}
T=\frac{1}{m}\sum_{\mu ,\nu =1}^{m}\sum_{ab}\left| t_{ab}^{\mu \nu }\right|
^{2}.  \label{g 3}
\end{equation}

We recall that it is for an e-PB interaction of the type described in Eq. (%
\ref{UND}) that

1) the {\it many-electron} Hamiltonian (with the e-e interaction switched
off) breaks up, in the D basis, into $m$ independent {\it single-electron}
Hamiltonians, and hence

2) the conduction problem breaks up in a similar manner and is soluble in
terms of {\it single-electron} quantities.

We also remind the reader that within our model the single-e-PB $S$ matrix,
and hence the transmission amplitudes appearing in Eq. (\ref{g 2}), are not
``generic'', but have a rather restricted structure.

The final answer, though, i.e. Eq. (\ref{g 2}), is a very appealing one and
is likely to be valid beyond the situation envisaged by the present model,
i.e. to cases where the e-PB $S$ matrix has a more general structure.
However, in such a generic case it is not known to us under what
approximations the conduction problem --even with the e-e Coulomb interaction
switched off-- could still be reduced to independent single-electron
problems and thus expressed in terms of single-electron quantities. The best
we can do at present is {\it conjecture} the validity of Eq. (\ref{g 2}) for
a generic e-PB $S$ matrix, under suitable approximations, as explained at
the end of section III around equation (\ref{lin}).

\section{A RANDOM-MATRIX MODEL FOR THE SCATTERING MATRIX}

\label{RMT}

In the past, quantum electronic transport in mesoscopic systems has been
described successfully in terms of ensembles of single-electron $S$ matrices 
(see, for instance, Refs. 
\onlinecite{baranger-mello95,markus model in cavities,les houches,wavesrm,baranger-mello94,carlo94,carlo97}). In particular, it was shown that, in the absence of direct
processes, {\it quantum transport through classically chaotic cavities} can
be studied in terms of the {\it invariant measure} for the $S$ matrix, which
is a precise mathematical formulation of the intuitive notion of
``equal-a-priori-probabilities'' in $S$-matrix space. For TRI systems, like
the ones we are studying here, the invariant measure is also known as the 
{\it Circular Orthogonal Ensemble }(COE) \cite{mehta}. The COE for $S$
matrices of dimensionality $2N$, $N$ being the number of open channels
supported by the two leads attached to the cavity, gives, for the ensemble
averaged (indicated by brackets $\left\langle \cdot \cdot \cdot
\right\rangle $) spinless conductance and its variance \cite
{wavesrm,baranger-mello94} 
\begin{equation}
\left\langle T\right\rangle =\frac{N^{2}}{2N+1},  \label{<T>}
\end{equation}
\begin{equation}
var(T)=\frac{N(N+1)^{2}}{\left( 2N+1\right) ^{2}\left( 2N+3\right) },
\label{varT}
\end{equation}
respectively. The $1$ in the denominator of Eq. (\ref{<T>}) is the
weak-localization correction (WLC).

We construct here an ensemble of $S$ matrices for the system consisting of a
single electron and the PB, and, using the conductance formula obtained in
the previous section, Eq. (\ref{g 2}), we analyze the effect of the PB on
the average and variance of the conductance.

Within the model we have been discussing for the e-PB system, we postulate $m
$ independent COE's for the $S$ matrices $\overline{S}^{\sigma }$ of Eq. (%
\ref{SD}). Ensemble averaging Eq. (\ref{g 1}) we obtain
\begin{eqnarray}
\left\langle T\right\rangle _{m} &=&\sum_{\sigma =1}^{m}p(\sigma
)\left\langle tr\left[ (\overline{t}^{\sigma })^{\dagger }\overline{t}%
^{\sigma }\right] \right\rangle   \nonumber \\
&=&\sum_{\sigma =1}^{m}p(\sigma )\frac{N^{2}}{2N+1}  \nonumber \\
&=&\frac{N^{2}}{2N+1},  \label{<T>m }
\end{eqnarray}
the same result as in Eq. (\ref{<T>}), in the absence of the PB. Thus the
model is not generic enough to decrease the WLC. For the variance we obtain 
\begin{eqnarray}
\left[ var(T)\right] _{m} &=&\sum_{\sigma =1}^{m}\left[ p(\sigma )\right]
^{2}var\left\{ tr\left[ (\overline{t}^{\sigma })^{\dagger }\overline{t}%
^{\sigma }\right] \right\}   \nonumber \\
&=&\frac{N(N+1)^{2}}{\left( 2N+1\right) ^{2}\left( 2N+3\right) }\sum_{\sigma
=1}^{m}\left[ p(\sigma )\right] ^{2}.  \label{varT m}
\end{eqnarray}
If one PB state $\sigma _{0}$ has probability 1 and all the others $0$, $%
\sum_{\sigma =1}^{m}\left[ p(\sigma )\right] ^{2}=1$. In the other extreme
case of equiprobable PB states, $\sum_{\sigma =1}^{m}\left[ p(\sigma
)\right] ^{2}=1/m$. The conductance thus fluctuates less than in the absence
of the PB, as expected \cite{baranger-mello95}.

Should the conjecture we made at the end of last section be true, and the
result (\ref{g 2}) be valid more generally, i.e. for a generic e-PB $S$
matrix, we could postulate a COE for the full $2mN$-dimensional $S$ matrix
in the ND basis. We would then obtain, ensemble averaging Eq. (\ref{g 2})
\begin{equation}
\left\langle T\right\rangle _{m}
=\sum_{\mu \nu \nu ^{\prime } } \sum_{ab} 
\rho _{PB}^{\nu \mu }\left\langle \left[ t_{ab}^{\nu ^{\prime }\mu }\right]
^{*}t_{ab}^{\nu ^{\prime }\nu }\right\rangle   \label{<T>m 1}
\end{equation}
From Ref. \onlinecite{wavesrm} we find 
\[
\left\langle \left[ t_{ab}^{\nu ^{\prime }\mu }\right] ^{*}t_{ab}^{\nu
^{\prime }\nu }\right\rangle =\frac{\delta _{\mu \nu }}{2mN+1}
\]
and, since $\sum_{\nu }\rho _{PB}^{\nu \nu }=1$, we finally obtain
\begin{equation}
\left\langle T\right\rangle _{m}=\frac{N^{2}}{2N+\frac{1}{m}}.
\label{<T>m 2}
\end{equation}
Now we observe that, as $m\rightarrow \infty $, the effect of the PB is to
kill the WLC, as expected. 

We find the variance of $T$ only in the simplified situation described by
Eq. (\ref{g 3})
\begin{equation}
\left[ var(T)\right] _{m}=\frac{1}{m^{2}}var\left[ \sum_{\mu ,\nu
=1}^{m}\sum_{ab}\left| t_{ab}^{\mu \nu }\right| ^{2}\right] .
\label{varT m 2}
\end{equation}
The variance appearing on the right hand side of this last equation can be
read form Eq. (\ref{varT}), with the replacement $N\Rightarrow mN$, to find
\begin{equation}
\left[ var(T)\right] _{m}=\frac{1}{m^{2}}\frac{N(N+\frac{1}{m})^{2}}{\left(
2N+\frac{1}{m}\right) ^{2}\left( 2N+\frac{3}{m}\right) }.  \label{varTm 3}
\end{equation}
For large $m$, conductance fluctuations are killed as well, as expected.

\section{CONCLUSIONS}
\label{concl}

We have discussed the quantum electronic conduction problem in a
mesoscopic system, allowing for the presence of a phase breaker (PB) the
system can interact with. The PB can exist in $m$ quantum-mechanical states.
It could represent an impurity with internal degrees of freedom, so that its
state may change via the interaction with the electrons: for instance, it
could be a magnetic impurity interacting with the electron spin. The PB\
might also represent the environment --for instance, the phonon field--
whose state is allowed to change.

We first studied the problem of a single electron interacting with the PB,
in order to investigate the effect of the latter in the interference terms
that are there in the absence of the PB. We described the static, as well as
the dynamical scatterers (the PB), in terms of their scattering or $S$
matrices: this makes the discussion very intuitive, quite general, and
amenable to the application of random-matrix models that have been developed
in the past. We found an $S$-matrix formulation that is capable of decribing
the crossover from a purely coherent response to a classical, or incoherent
one.

We then set out to study the conduction problem --a many-electron problem.
The e-e interaction is treated in an RPA approximation. We could make the very
general statement (that so far, to our knowledge, was known only in the
absence of a PB) that in the dc limit one can give an explicit
expression for the current in terms of the potential difference applied
between the two reservoirs, the full potential profile not being needed.
That expression involves the conductivity tensor $\sigma _{\alpha \beta
}^{0}({\bf r},{\bf r}^{\prime })$ which, in the absence of the PB, can be
calculated in terms of single-particle Green's function and, eventually,
single-particle transmission coefficients. In the presence of the PB, the
e-e interaction induced by the PB has not allowed us to follow a
similar path. That induced interaction is there even in the absence of the 
e-e Coulomb interaction, which, within RPA, was disposed of and replaced, in turn, by
the solution of a self-consistent problem.

We proposed a model for the e-PB interaction that can be diagonalized and
transformed into $m$ single-particle problems. 
The conduction problem splits into $m$ single-particle
problems as well, and the final result for the conductance, given in 
Eq. (\ref{g 2}), is like the standard one without a
PB, except that now an extra trace over the $m$ PB states appears. We should
stress that, within the model, result (\ref{g 2}) is {\it not perturbative}
in the e-PB interaction strength. However, the model
implies a single-electron-PB $S$ matrix of a rather restricted form. For a
``generic'' single-electron-PB $S$ matrix we do not even know under what approximations
(that would imply disregarding the induced e-e interaction) the conduction problem can
be reduced to a single-electron one. The result expressed in 
Eq. (\ref{g 2}) is so intuitive, though, that we conjecture its validity under a more
general e-PB $S$ matrix, within some suitable approximation, as discussed  
at the end of section III around equation (\ref{lin}).

It would be instructive to solve this same conduction problem within the
spirit of Landauer's approach \cite{landauer(phil mag)}, just as in Ref. 
\onlinecite{chen-sorbello}, and verify that one arrives at the result (\ref{g 2})
under the special model used here and not in general. A suitable
approximation for treating the more general problem might suggest itself in
that approach. But this we have not succeeded to do so far.

Finally, a random-matrix model was setup for the description of the e-PB
system, with possible applications to chaotic cavities: the effect of the PB
on the conductance average and its fluctuations was analyzed. The
limitations of the model became apparent in that study; in contrast, a
generic e-PB $S$ matrix was shown to give much more freedom in the
description of the weak-localization correction and the conductance
fluctuations.

\section*{acknowledgements}

One of the authors (PAM) acknowledges partial financial support from
CONACyT, Mexico, through Contract No. 2645P-E, as well as the hospitality 
and partial financial support of the I.T.P. of the Technion, Haifa, 
and of the Weizmann Institute, Rehovot, where parts of
this work were discussed. He also acknowledges fruitful discussions with H.
Su\'{a}rez and with M. B\"{u}ttiker, the latter during a stay at the
Universit\'{e} de Gen\`{e}ve, whose hospitality is greatly acknowledged. We 
acknowledge instructive discussions with the late R. Landauer and with
D. E. Khmelnitskii. The
research was partially supported by the DIP project on ``Quantum Electronics
in Low-Dimensional Systems'' and by the GIF project on ``Electron Interactions 
and Disorder in Finite Conductors".

\end{multicols}

\end{document}